\documentclass[12pt]{article}

\pdfoutput=1

\usepackage[english]{babel}
\usepackage{fancyhdr}
\usepackage{amsmath}
\usepackage{amssymb}
\usepackage{amsfonts}
\usepackage{psfrag}
\usepackage[applemac]{inputenc}
\usepackage{graphicx,wrapfig}
\usepackage{pstricks}

\usepackage{slashed}            
\usepackage{subfig}             
\usepackage{xspace}				
\usepackage[sort&compress,square,comma,numbers]{natbib}
\usepackage{float}
\usepackage{dcolumn}
\usepackage{bm}
\usepackage{epstopdf}
\usepackage{feynmp}

\usepackage[
	colorlinks=true,
	citecolor=black,
	linkcolor=black,
	urlcolor=blue,
	hypertexnames=false]{hyperref}
	
\newcommand{\arXiv}[2]{\href{http://arxiv.org/pdf/#1}{{\tt #2/#1}}}
\newcommand{\arXivold}[1]{\href{http://arxiv.org/pdf/#1}{{\tt #1}}}

\DeclareFontFamily{OT1}{pzc}{}
\DeclareFontShape{OT1}{pzc}{m}{it}{<-> s * [1.100] pzcmi7t}{}
\DeclareMathAlphabet{\mathpzc}{OT1}{pzc}{m}{it}

\newcommand{\beq}{\begin{eqnarray}}
\newcommand{\eeq}{\end{eqnarray}}
\newcommand{\bea}{\begin{eqnarray}}
\newcommand{\eea}{\end{eqnarray}}
\newcommand{\bag}{\begin{align}}
\newcommand{\eag}{\end{align}}

\newcommand{\TeV}{\,\mathrm{TeV}}

\begin{document}

\baselineskip=18pt

\setcounter{footnote}{0}
\setcounter{figure}{0}
\setcounter{table}{0}


\begin{titlepage}

\begin{center}
  \begin{LARGE}
    \begin{bf}
The Minimal Model of a Diphoton Resonance: 

\vspace*{0.2cm}

Production without Gluon Couplings
   \end{bf}
  \end{LARGE}
\end{center}
\vspace{0.1cm}
\begin{center}
\begin{large}
{\bf Csaba Cs\'aki$^a$, Jay Hubisz$^b$, John Terning$^c$ \\}
\end{large}
  \vspace{0.5cm}
  \begin{it}

\begin{small}
$^{(a)}$Department of Physics, LEPP, Cornell University, Ithaca, NY 14853, USA
\vspace{0.2cm}\\
$^{(b)}$Department of Physics, Syracuse University, Syracuse, NY 13244, USA
\vspace{0.2cm}\\

$^{(c)}$Department of Physics, University of California, Davis, CA 95616
 \vspace{0.1cm}

\end{small}

\end{it}
\vspace{.5cm}

{\tt csaki@cornell.edu, jhubisz@syr.edu, jterning@gmail.com}

\end{center}

\vspace*{0.5cm}

\begin{abstract}
\medskip
\noindent
We consider the phenomenology of a resonance that couples to photons but not gluons, and estimate its production rate at the LHC from photon-photon fusion in elastic pp scattering using the equivalent photon and narrow width approximations.  The rate is sensitive only to the mass, the spin, the total width of the resonance, and its branching fraction to photons.  Production cross sections of 3-6 fb at 13 TeV can be easily accommodated for a 750 GeV resonance with partial photon width of 15 GeV. This provides the minimal explanation of the reported diphoton anomaly in the early LHC Run II data.

\end{abstract}

\bigskip

\end{titlepage}


The ATLAS and CMS experiments have recently reported excesses \cite{data} in the diphoton invariant mass spectrum of 13 TeV proton-proton collisions at the LHC. A relatively broad peak with a width of $\Gamma \sim 20-45$ GeV has been observed at the mass $m\sim 750$ GeV, while the number of events would correspond to a production rate of $\sigma (pp\to R+X) {\rm Br} (R\to \gamma\gamma ) \sim 3-6$ fb.  If confirmed this phenomenon would provide the first direct collider physics evidence for physics beyond the standard model.  While the current excesses (3.8 $\sigma$ local at ATLAS and 2.6 $\sigma$ local at CMS) are not sufficient to draw a definite conclusion, this hint is nevertheless very intriguing and provides motivation to understand all possible explanations that would result in such a signal without already being excluded from the various searches \cite{Aad:2015mna,CMSgg8} at Run I of the LHC at 7 and 8 TeV. Most existing explanations \cite{resonaances,Harigaya:2015ezk,Mambrini:2015wyu,Backovic:2015fnp,Angelescu:2015uiz,Nakai:2015ptz,Knapen:2015dap,Buttazzo:2015txu,Pilaftsis:2015ycr,Franceschini:2015kwy,DiChiara:2015vdm,Higaki,McDermott,Ellis,Low,Bellazzini,Gupta,Petersson:2015mkr,Molinaro:2015cwg}
 focus on a scalar resonance (like a heavy Higgs boson) produced via gluon fusion, with a subsequent decay to two photons, in complete analogy to one of the two main original discovery channels of the SM Higgs at 125 GeV. However the fact that this resonance appears to be much broader than the Higgs boson suggests that its production and decay might actually be quite different from that of the ordinary Higgs. Since no other decay channel has been observed, there should be a significant branching fraction into photons, the only channel where an excess has been seen to date. This implies that the coupling to photons should be quite sizable. 

Once the photon branching fraction is ${\cal O} (1)$, a new possibility of producing such a resonance arises: photon-photon fusion in elastic pp scattering \cite{Jaeckel:2012yz}. The reason for this process being negligible ($\sigma(pp\to p p \gamma\gamma\to pp h)\sim$ 0.1 fb \cite{Khoze:2001xm}) for Higgs production is that the Higgs branching fraction to photons is very small. Once the coupling to photons is sizable, as with this conjectured resonance, a cross section of order 10 fb can be easily achieved.

The purpose of this paper is to demonstrate that the diphoton excess can be explained by only coupling the resonance to photons, and considering elastic scattering processes with photon fusion into the resonance without a disassociation of the protons, which provides a lower bound on the total cross section. This process can have a sufficiently large production cross section, and would explain why no other associated objects are observed in the events. It would also give a plausible explanation for the absence of any signals in other channels, including dijet resonances. Depending on the exact value of the parameter appearing in the photon parton distribution function (PDF) the ratio of production rates for Run I and Run II energies varies between 6 and 10. 


\vspace*{1cm}

We will be considering a model with an additional scalar particle $R$ of mass $m \approx 750$ GeV whose only sizeable coupling to SM particles is to photons via the operator 
\begin{equation}
\frac{c_{\gamma\gamma}}{v} R F^2
\end{equation}
The resulting partial width to photons $\Gamma_{\gamma\gamma}$ is 
\begin{equation}
\Gamma_{\gamma\gamma} = \frac{c_{\gamma\gamma}^2}{4\pi} \frac{m^3}{v^2}\ .
\label{eq:width}
\end{equation}

The production process we will consider is the coherent emission of a photon from each proton and the subsequent production of the resonance by the two photons, as shown in Fig.~1. In this process the protons are elastically scattered at very small angles and the photons are almost on shell.

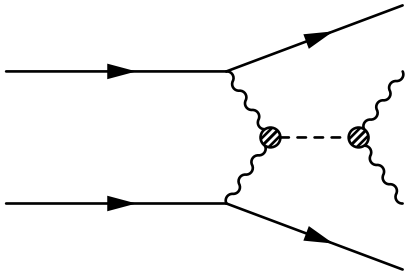
\begin{figure}
\begin{center}
	    \begin{fmffile}{gammagammascattering}
	        \begin{fmfgraph*}(150,100)
                 \fmfstraight
	        \fmfleft{i1,i2,i3,i4,i5}
	        \fmfright{o1,o2,o3,o4,o5}
                 \fmf{fermion}{i2,v1,o1}
                  \fmf{fermion}{i4,v3,o5}
                  \fmf{wiggly,label=$\gamma$}{v1,v2}
                  \fmf{wiggly,label=$\gamma$}{v3,v2}
                  \fmf{dashes,label=$R$}{v2,v4}
\fmflabel{$p$}{i2}
\fmflabel{$p$}{i4}
\fmflabel{$p$}{o1}
\fmflabel{$p$}{o5}
\fmflabel{$\gamma$}{o2}
\fmflabel{$\gamma$}{o4}
                  \fmf{wiggly}{v4,o4}
                  \fmf{wiggly}{v4,o2}
\fmfblob{0.05w}{v2,v4}

	        \end{fmfgraph*}
	    \end{fmffile} 
	    \caption{Resonance production via photon-photon fusion in elastic p-p scattering\label{Fig:elastic}}
\end{center}
\end{figure}

The cross section for a general $A+B \rightarrow R \rightarrow C+D$ hard scattering process with no incoming color multiplicities is given by
\begin{equation}
\sigma(\hat{s}) = 32 \pi \frac{2 J_R + 1}{(2 J_A + 1) (2 J_B + 1)} \frac{ \Gamma_{AB} \Gamma_{CD} }{(\hat{s} - m^2)^2+ m^2 \Gamma^2}
\end{equation}
where the $(2J+1)$ factors are spin multiplicities of the initial and intermediate states (these are each 2 for the incoming photons), and $\hat{s}$ is the invariant mass of the particles $A,B$ (the two fusing photons in our case).  Using the narrow width approximation the phase space for the cross section is simplified:
\begin{equation}
\frac{d\hat{s}}{(\hat{s} - m^2)^2+ m^2 \Gamma^2} \rightarrow \frac{\pi}{m \Gamma} \delta (\hat{s} - m^2)
\end{equation}
Using this approximation and applying it to proton-proton scattering we can simplify the total cross section involving the convolution of PDFs with the $\hat{s}$-dependent cross section.  

For our process of interest, we apply the equivalent photon approximation, where the photon has a distribution function at $E_\text{CM}^2 = s$ given by $f^\gamma_s (x)$, where $E_\text{CM}$ is the center of mass energy of the proton-proton collision, and $x$ is the fraction of the proton energy carried by the photon:
\begin{equation}
\sigma(pp \to pp R) = \int dx_1 dx_2 f^\gamma_s (x_1) f^\gamma_s (x_2) \sigma(x_1 x_2 s)
\end{equation}
Note that the energy of the photon-photon system is given by $\hat{s} = x_1 x_2 s$.

The cross section is then given by
\begin{equation}
\sigma(pp \to pp \gamma\gamma) = \frac{8 \pi^2}{m} \frac{\Gamma_{\gamma\gamma}^2}{\Gamma} (2 J_R+1) \int dx_1 dx_2 f^\gamma_s (x_1) f^\gamma_s (x_2) \delta( x_1 x_2 s - m^2 )
\end{equation}
Eliminating the delta function, we have
\begin{equation}\sigma(pp \to pp \gamma\gamma) = \frac{8 \pi^2 \Gamma}{s m} \text{Br}^2(R\rightarrow \gamma\gamma)  (2 J_R+1) \int \frac{dx}{x} f^\gamma_s \left(x \right) f^\gamma_s \left(\frac{m^2}{x s}\right) 
\end{equation}

 The equivalent photon approximation \cite{Weizsacker} gives the dominant contribution to the photon distribution function in two photon production which has the following approximate form for small $x$ \cite{Budnev:1974de,Martin:2014nqa}:
 \begin{equation}
f_s^\gamma(x) dx = \frac{dx}{x} \frac{2 \alpha}{\pi} \log \left[ \frac{q_*}{m_p} \frac{1}{x} \right].
\end{equation}
Here $q_*$ is the inverse of the minimum impact parameter for elastic scattering which is naively set by the proton radius:
\beq
\frac{1}{q_*}=r_p = 0.8768 \,\,{\rm fm} = \frac{1}{170 \,{\rm MeV}}~.
\eeq
Hereafter, we use the replacement $r_* = q_*/m_p$, and $r_m = m/\sqrt{s}$

A useful cross check on the value of $q_*$ is that plugging the Higgs mass and partial width into our formulae reproduces the the LHC two photon Higgs production cross section
($\sigma = 0.1$ fb \cite{Khoze:2001xm}) for 130 MeV $< q_* <$ 170 MeV. The photon PDF calculation of ref.~\cite{Martin:2014nqa} which gives the momentum fraction of the proton carried by the photon at $Q^2= 1$ GeV$^2$ also implies an upper bound: $q_* < 170$ MeV. 

\begin{figure}[htb]
\begin{center}
\includegraphics[width=10cm]{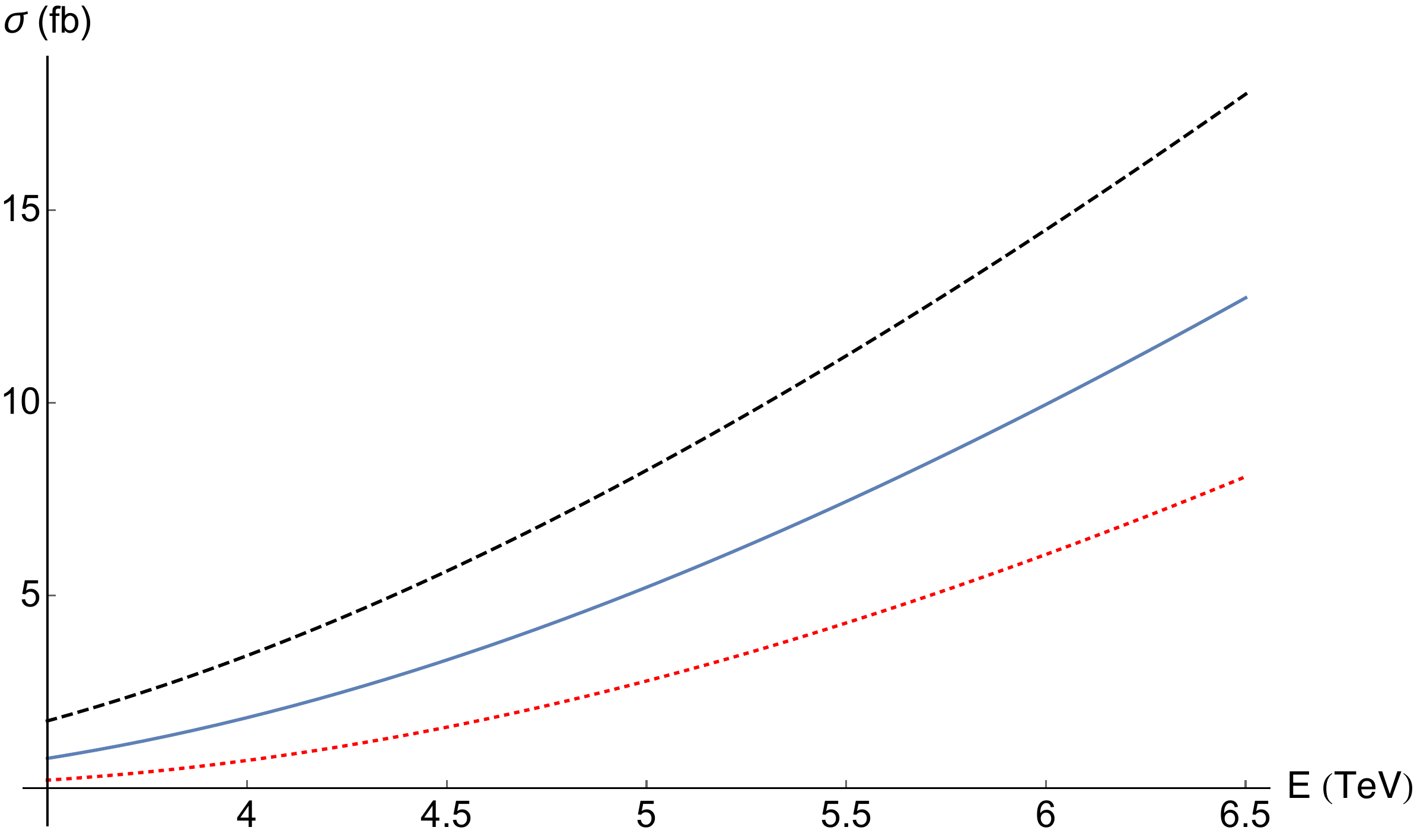}
\end{center}
\caption{The the two photon production cross section as a function of beam energy with a partial width $\Gamma_{\gamma\gamma}=15$ GeV, total width $\Gamma=45$ GeV, and three values of $q_*$, 130 (dotted) 150 MeV (solid) and 170 MeV (dashed).} \label{fig:crosssection}
\end{figure}

Substituting the distribution function into the cross section formula, we get
\begin{equation}
\sigma(pp \to pp \gamma\gamma) = \frac{32 \alpha^2\Gamma}{m^3} \text{Br}^2(R\rightarrow \gamma\gamma)  (2 J_R+1) \int_{x_\text{min}}^{x_\text{max}} \frac{dx}{x} \log (r_*/x) \log (x r_*/r_m^2)
\end{equation}
The requirement of elasticity gives $x_\text{max} = r_*$, and thus a corresponding $x_\text{min} = r_m^2/r_*$.  
Note that for $\gamma r_* m_p=\gamma q_* = m/2$ we have $x_{\rm min}=x_{\rm max}$ and the elastic two photon production cross section is zero. The physics behind this is very simple, since $\gamma q_*$ is the maximum photon energy, we cannot produce the resonance at all if $\gamma q_* $ is less than $m/2$.
For larger values of $x$, the proton begins to dissociate, and the photon distribution function then drops off steeply.  For the same reason, the production via $Z$ fusion which would naively be expected to contribute is small, as radiation of an on-shell $Z$ must necessarily dissociate the proton due to the large momentum transfer required.
The integral over the distribution functions is in fact analytic, and the result is 
\begin{equation}
\int_{x_\text{min}}^{x_\text{max}} \frac{dx}{x} \log (r_*/x) \log ( x r_* /r_m^2) = \frac{4}{3} \log^3 (r_*/r_m).
\end{equation}
The final cross section for the process $pp \rightarrow R \rightarrow \gamma\gamma$ is then given by
\begin{equation}
\sigma(pp \to pp \gamma\gamma\to pp R) = \frac{128 \alpha^2\Gamma}{3 m^3} \text{Br}^2(R \rightarrow \gamma\gamma) (2 J_R+1)  \log^3 \left(\frac{r_*}{r_m}\right)
\end{equation}
We show the resulting cross section in Fig.~\ref{fig:crosssection} as a function of beam energy.

\begin{figure}[htb]
\begin{center}
\includegraphics[width=10cm]{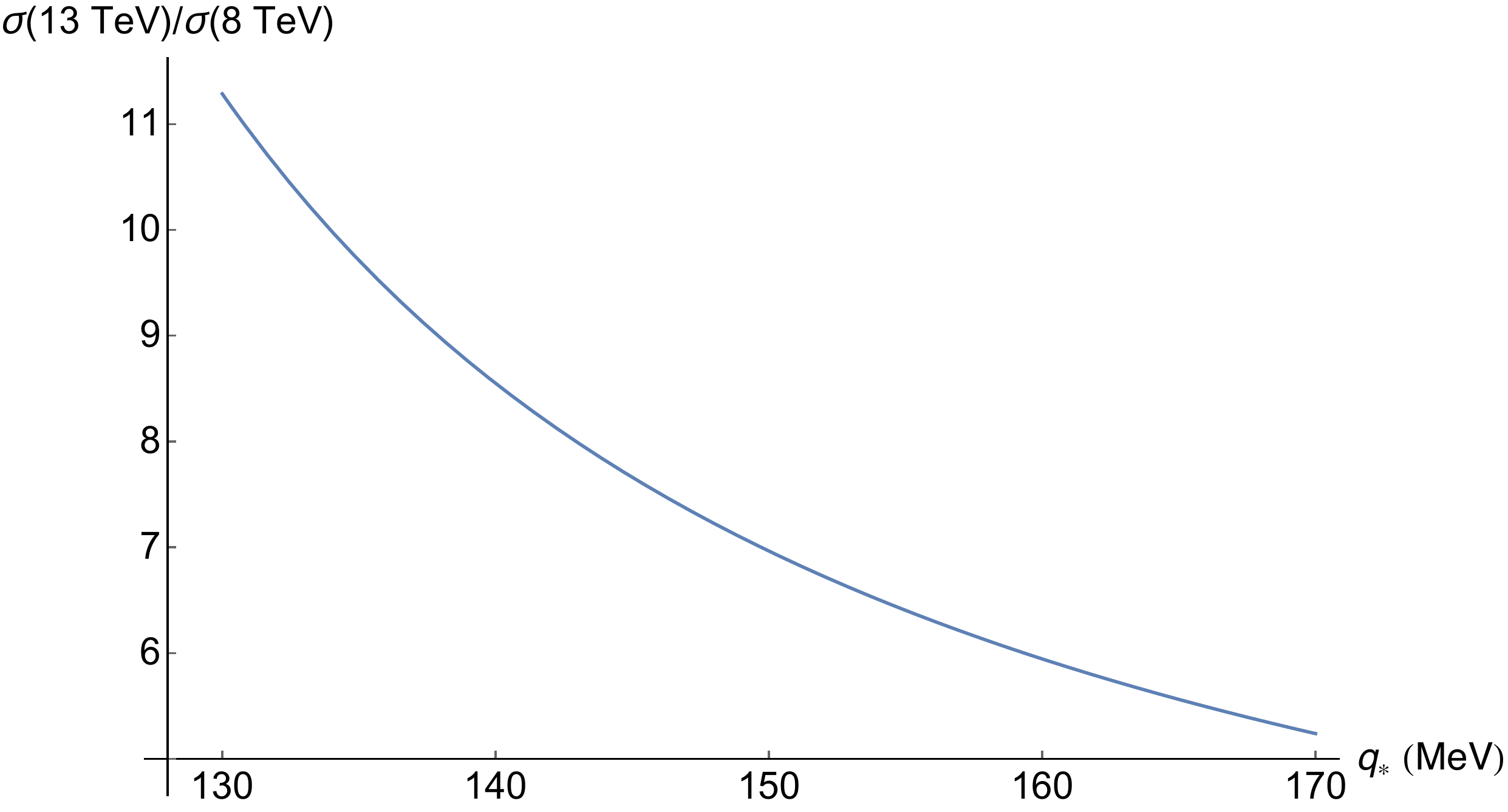}
\end{center}
\caption{The ratio of the two photon production cross section at 13 and 8 TeV as a function of $q_*$.} \label{fig:ratio}
\end{figure}

For a spin zero resonance with mass $750$~GeV, and taking $q_* = 170$~MeV, we have the following results for the total near-resonance cross section at $8$~TeV and $13$~TeV:
\begin{align}
\sigma_{8~\text{TeV}} &= 31~\text{fb}~\left(\frac{\Gamma}{45~\text{GeV}} \right) \text{Br}^2(R \rightarrow \gamma\gamma) \\
\sigma_{13~\text{TeV}} &= 162~\text{fb}~\left(\frac{\Gamma}{45~\text{GeV}} \right) \text{Br}^2(R \rightarrow \gamma\gamma).
\end{align}
For a value of $q_*=130$~MeV on the lower end of the window suggested by comparison with Higgs data, we have
\begin{align}
\sigma_{8~\text{TeV}} &= 6.5~\text{fb}~\left(\frac{\Gamma}{45~\text{GeV}} \right) \text{Br}^2(R \rightarrow \gamma\gamma) \\
\sigma_{13~\text{TeV}} &= 73~\text{fb}~\left(\frac{\Gamma}{45~\text{GeV}} \right) \text{Br}^2(R \rightarrow \gamma\gamma).
\end{align}
In Fig.~\ref{fig:ratio} we show the ratio of cross sections at the two energies as a function of $q_*$. We see that smaller values of $q_*$ correspond to a much larger increase in going from 8 TeV to 13 TeV.

\vspace*{1cm}

We have seen that a successful explanation of the diphoton excess requires a relatively large partial photon width $\Gamma_{\gamma\gamma} \sim 15$ GeV, which using (\ref{eq:width}) implies $c_{\gamma\gamma} \sim 0.16$. Loops of weakly interacting weak scale particles would however result in much smaller photon couplings. An NDA estimate would be 
\beq
c_{\gamma\gamma} \sim \frac{\alpha}{8\pi} \frac{v}{M}~,
\eeq
 where $M$ is the mass of the heavy particle running in the loop responsible for the generation of this coupling. We can see that the right magnitude would be generated if $c_{\gamma\gamma}$ was the result of a strongly coupled loop with $\alpha_{\rm eff} \sim 4\pi$ and $v/M \sim 0.1$ implying strong dynamics at the multi-TeV scale. 

\vspace*{1cm}

We have shown that elastic proton scattering with two photon fusion  can provide a large enough cross section  to produce  the 750 GeV resonance provided that the $\gamma \gamma $ partial width is around 1/3 to 1/2 of the total width. This eliminates possible deviations in dijet spectra and opens up new avenues for model building that do not include new colored particles. If two photon production is the main production process then the events should be mostly central with large rapidity gaps.

\vspace*{1cm}
{\bf Note Added:} After submitting this paper we became aware of ref. \cite{Fichet:2015vvy} which also considers two photon production, including even larger inelastic two photon processes as well.

\section*{Acknowledgments}

We thank Markus Luty and Ennio Salvione for useful discussions. C.C.~is supported in part by the NSF grant PHY-1316222.  J.H.~is supported in part by the DOE under grant DE-FG02-85ER40237.  J.T.~is supported in part by the DOE under grant DE-SC-000999.



\end{document}